\documentclass[journal,twocolumn]{IEEEtran}
\usepackage{rotating}
\usepackage{graphicx}
\usepackage{epsfig}
\usepackage{cite}
\usepackage{color}
\usepackage{units}

\begin{document}
\title{Progress and Validation of Geant4 Based Radioactive Decay
  Simulation Using the Examples of Simbol-X and IXO}

\author{S.~Hauf,~M.~Kuster,~M.G.~Pia,~Z.~Bell,
  ~U.~Briel,~R.~Chipaux,~D.H.H. Hoffmann,
  ~E.~Kendziorra,~P.~Laurent,~L.~Str\"uder,~C.~Tenzer,~G.~Weidenspointner,~A.~Zoglauer
\thanks{Manuscript received November 20, 2009.}
\thanks{S. Hauf, M. Kuster and D.H.H. Hoffmann are with TU Darmstadt, Institut f\"ur 
Kernphysik, Schlossgartenstrasse 9, D-64289 Darmstadt, Germany}
\thanks{M.G. Pia is with European Organization for Nuclear Research
(CERN), CH-1211 Gen\`eve 23, Switzerland and INFN, Sezione di Genova,
Genova, Italy}
\thanks{Z. Bell is with Oak Ridge National Laboratry, Oak Ridge, TN, USA}
\thanks{U. Briel is with Max-Planck-Institut f\"ur extraterrestrische
Physik, Giessenbachstrasse, D-85748 Garching, Germany}
\thanks{R. Chipaux and P. Laurent are with CEA/DSM/IRFU, Centre de Saclay,
F-91191 Gif sur Yvette, France}
\thanks{E. Kendziorra and C. Tenzer are with Institut f\"ur Astronomie und
Astrophysik T\"ubingen, Sand 1, D-72076 T\"ubingen}
\thanks{L. Str\"uder is with MPI Halbleiterlabor, Otto-Hahn-Ring 6,
D-81739, Munich, Germany and Max-Planck-Institut f\"ur extraterrestrische
Physik, Giessenbachstrasse, D-85748 Garching, Germany}
\thanks{G. Weidenspointner is with MPI Halbleiterlabor, Otto-Hahn-Ring 6,
D-81739, Munich, Germany}
\thanks{A. Zoglauer is with University of California, Space Science
Laboratory, Berkeley, USA}
}

%% \author{\IEEEauthorblockN{S. Hauf\IEEEauthorrefmark{1},
%%     M. Kuster\IEEEauthorrefmark{1}, M.G. Pia\IEEEauthorrefmark{2,3},
%%     Z. Bell\IEEEauthorrefmark{4}, U. Briel \IEEEauthorrefmark{6},
%%     R. Chipaux \IEEEauthorrefmark{8},
%%     D.H.H. Hoffmann\IEEEauthorrefmark{1},\\ E. Kendziorra\IEEEauthorrefmark{9},
%%     P. Laurent \IEEEauthorrefmark{8},
%%     L. Str\"uder\IEEEauthorrefmark{5,6}, C. Tenzer
%%     \IEEEauthorrefmark{9}, G. Weidenspointner \IEEEauthorrefmark{5},
%%     A. Zoglauer \IEEEauthorrefmark{7}}
%%   \IEEEauthorblockA{\IEEEauthorrefmark{1}Technische Universit\"at Darmstadt, Institut f\"ur Kernphysik, Schlossgartenstrasse 9,\\ D-64289 Darmstadt, Germany}
%%   \IEEEauthorblockA{\IEEEauthorrefmark{2}European Organization for Nuclear Research (CERN), CH-1211 Gen\`eve 23, Switzerland}
%%   \IEEEauthorblockA{\IEEEauthorrefmark{3}INFN, Sezione di Genova, Genova, Italy}
%%   \IEEEauthorblockA{\IEEEauthorrefmark{4}Oak Ridge National Laboratory}
%%   \IEEEauthorblockA{\IEEEauthorrefmark{5}MPI Halbleiterlabor, Otto-Hahn-Ring 6, D-81739 M\"unchen, Germany}
%%   \IEEEauthorblockA{\IEEEauthorrefmark{6}Max-Planck-Institut f\"ur extraterrestrische Physik, Giessenbachstrasse, D-85748 Garching, Germany}
%%   \IEEEauthorblockA{\IEEEauthorrefmark{7}University of California, Space Sience Laboratory, Berkeley, USA}
%%   \IEEEauthorblockA{\IEEEauthorrefmark{8}CEA/DSM/IRFU, Centre de Saclay, Gif-sur-Yvette, France}
%%   \IEEEauthorblockA{\IEEEauthorrefmark{9}Institut f\"ur Astronomie und Astrophysik T\"ubingen, Sand 1, D-72076 T\"ubingen, Germany}
%% } 

\maketitle
\pagestyle{empty}
\thispagestyle{empty}

\begin{abstract} \boldmath
The anticipated high sensitivity and the science goals of the next
generation X-ray space missions, like the International X-ray
Observatory or Simbol-X, rely on a low instrumental background, which
in turn requires optimized shielding concepts. We present Geant4 based
simulation results on the IXO Wide Field Imager cosmic ray proton
induced background in comparison with previous results obtained for
the Simbol-X LED and HED focal plane detectors. Our results show that
an improvement in mean differential background flux compared to
actually operating X-ray observatories may be feasible with detectors
based on DEPFET technology. In addition we present preliminary results
concerning the validation of Geant4 based radioactive decay simulation
in space applications as a part of the Nano5 project.
\end{abstract}

\section{Introduction}
\IEEEPARstart{T}{he} next generation X-ray space missions like the
International X-ray Observatory IXO, Simbol-X, NuStar or Astro-H aim
to explore the X-ray sky in the energy range between 0.1 and 80 keV
with so far unrivalled high sensitivity ~\cite{NuStar,
  takahashi-2008}. To achieve this goal both missions require a low
instrumental background which can only be realized with optimized
shielding and background reduction techniques. To optimize the trade
off between cost, weight, and performance of the detectors and
shielding components, extensive and reliable Monte-Carlo simulations
are necessary. Most of the state-of-the-art approaches to estimate the
prompt cosmic rays, solar proton and the cosmic X-ray induced
background in space rely on simulations with the Geant4 Monte Carlo
tool-kit~\cite{geant4:phys,GEANT4:2006}. The Geant4 electromagnetic
and hadronic physics models have extensively been verified not only
with space but also with ground based experiments. In contrast
measurements to verify the radioactive decay implementation in Geant4
have been rare or have only been tested on a limited set of isotopes,
which are not necessarily those used in satellite construction. On the
other hand, measured background data of actual and past missions
(e.g. INTEGRAL) show that up to 20\% of the instrumental background
can be due to long term activation of the detector materials in
orbit~\cite{jean:97a,jean:03}. This necessitates that the delayed
background component is also taken into account, well understood and
verified with laboratory measurements. While the background estimates
for Simbol-X and IXO presented in this work are focused on the prompt
cosmic ray proton induced background and optimizing the detector
shielding against resulting secondary particles, we also present a
first comparision of the radioactive decay physics implementation in
Geant4 with experimental measurements.

\begin{table}[!t]
  \renewcommand{\arraystretch}{1.3} \centering
  \caption{Simbol-X and IXO mission parameters.\label{tab:simbxIXO}}
  \begin{tabular}{lrr}
    \hline
                       & \multicolumn{1}{c}{Simbol-X}          & \multicolumn{1}{c}{IXO}\\ \hline
    Concept            & Formation flight                      & Expandable Bench\\ 
    Focal Length       & $25\,\mathrm{m}$                      & $20\,\mathrm{m}$ \\ 
    Energy Ranges      & $0.5$--$20\,\mathrm{keV}$             & $0.1$--$15\,\mathrm{keV}$ \\
                       & $5$--$80\,\mathrm{keV}$               & $5$--$40\,\mathrm{keV}$ \\ 
    Spatial Resolution & $128\times128\,\mathrm{pixels}$       & $1024\times 1024\,\mathrm{pixels}$ \\ 
    Angular Resolution & $30"$                                   & $5"$ \\ 
    Pixel Size         & $625\,\mathrm{\mu m}$                 & $100\,\mathrm{\mu m}$ \\ 
    Readout rate       & $4000\,\mathrm{Hz}$                   & $400\,\mathrm{Hz}$ \\ \hline
    %Background Reduction Methods & Anti-coincidence system, Graded-Z shielding, pattern and MIP detections & Graded-Z shielding, pattern and MIP detections \\ \hline
  \end{tabular}
\end{table}

\section{The Simbol-X Low and High Energy Detector}
The Simbol-X spacecraft is a planned X-ray observatory sensitive in
the energy range between 0.1 to 80
keV~\cite{ferrando:03a,ferrando:05a}. Focusing X-rays up to this
energy range requires a focal length of around
$20\,\mathrm{m}$. Because a satellite this large would be problematic
to launch with available launch systems, Simbol-X will consist of two
spacecrafts in close formation flight. The Simbol-X focal plane
consists of two detectors which cover the full energy range with a
maximum possible sensitivity.

The Simbol-X Low Energy Detector (LED) is a $450\,\mu\mathrm{m}$
thick, fully depleted DEPFET macro-pixel detector sensitive in the
energy range of $0.5$--$20\,\mathrm{keV}$~ \cite{lechner:03a}. The
current detector design provides an energy resolution of $E/\Delta E =
40$--$50$ at $6$--$10\,\mathrm{keV}$ \cite{ferrando:02a}. The major
advantages of this monolithic devices are there homogeneous entrance
window, a filling factor of $100\%$, the fast read-out and a quantum
efficiency above $98\%$ between $1$ and
$10\,\mathrm{keV}$. Furthermore, the DEPFET concept allows to reduce
the power consumption of the detector to a necessary minimum, since
the amplifiers of the individual pixels need only be powered during
read-out. The detector area is also homogeneously transparent which
allows for placing detectors sensitive in higher energy ranges
underneath. In the case of Simbol-X this is a CdTe High Energy
Detector (HED) sensitive in the $5$--$80$ keV
range~\cite{ferrando:02a}. The Simbol-X LED detector is subdivided
into $128\times 128$ pixels with a size of
$625\times625\,\mu\mathrm{m}^2$ providing an angular resolution of
$30$ arc seconds oversampling the mirror resolution by a factor of
$3$. The smallness of the LED detector allows a high read-out rate of
$8000\,\mathrm{Hz}$ making it possible to combine the detector with an
active anti-coincidence system, reducing the particle induced
background by an order of magnitude~\cite{tenzer:09a}. To suppress
secondary X-rays in the detector energy range of interest, induced by
particle and gamma-ray interactions in the detector materials, the
focal plane detector assembly is surrounded by a graded-Z shield
consisting of layers of tantalum, tin, copper, aluminum and a
carbon-composite material. For simulations the LED and HED along with
their surrounding shielding, the anti-coincidence and support
structures were modelled. The satellite structure and auxiliary
systems were replaced by a bulk aluminum mass with an expected mean
density. Data post-processing included proper anti-coincidence
treatment, as well as pattern and MIP analysis similar to the pattern
recognition and MIP rejection algorithm actually implemented in the
EPIC pn-camera event analyzer on board of XMM-Newton~\cite{hauf:09a}.

\section{The Wide Field Imager and Hard X-ray Imager of IXO}
The International X-ray Observatory (IXO) is a joint project of the
space agencies ESA, NASA and JAXA with the goal to develop a next
generation low background X-ray telescope with high resolution imaging
and spectroscopic observations up to $40\,\mathrm{keV}$. The
combination of these performance parameters requires a large effective
telescope area in combination with a low instrumental background. The
core IXO imaging instrument for the $0.1$--$15\,\mathrm{keV}$ energy
range will be the Wide Field Imager (WFI). The WFI concept follows a
similar design as the Simbol-X LED, i.e. silicon drift detector
macro-pixels with a DEPFET read-out. The detector consists of a
$1024\times 1024$ pixel array with $100\times 100\mu\mathrm{m}^2$
pixel size. The better mirror resolution in combination with the
smaller pixel size of the WFI as compared to Simbol-X leads to an
angular resolution of $3$ arc seconds in the energy range of
$0.1$--$15 \mathrm{keV}$ \cite{stefanescou:09a2}. The larger amount of
pixels will result in a higher data rate, which reduces the feasible
read-out rate due to power, telemetry and on board data handling
capacity from $8000$ to $400 \mathrm{Hz}$. At this read-out speed an
active anti-coincidence is rather unrealistic since it would result in
a dead time beyond 50\%. Similar to the Simbol-X LED design it is
planned to implement a graded-Z shield which is currently being
optimized in the course of our simulations.

The IXO Hard X-ray Imager (HXI) will cover photon energies up to
$40\,\mathrm{keV}$ with a $1024\times 1024$ pixel CdTe detector
array. It takes advantage of the fact that the WFI is homogeneously
transparent in this energy range. It will have the same pixel geometry
and spatial resolution as the WFI allowing for simultaneous spectra
imaging using both detectors. A comparison between the satellite
concepts of IXO and Simbol-X can be found in Table~\ref{tab:simbxIXO}.
\begin{figure}
  \centering
  \includegraphics[width=0.99\columnwidth]{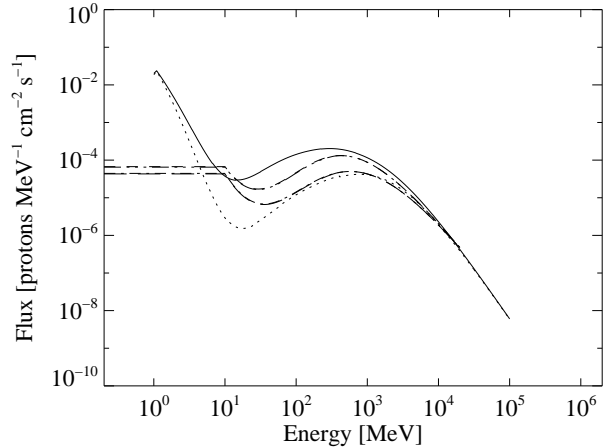}
  \caption{Spectral distribution of cosmic ray protons based on the
    CREME96 and CREME86 space radiation models for different mission
    and orbital parameters. From top to bottom (at 100 MeV) proton
    spectra assuming the following models and launch dates are shown:
    CREME96 for the IXO orbit assuming 2020 as launch date, CREME86
    for IXO orbit assuming 2020 as launch date, CREME86 for the IXO
    orbit assuming 2013 as launch date and CREME96 for the IXO orbit
    assuming 2013 as launch date. The two CREME86 Simbol-X orbit
    models are not distinguishable from their IXO counterparts.
    \label{fig:cremeComp}}
\end{figure}

\section{The Cosmic Ray Spectrum at L2}
The IXO spacecraft will be positioned at the L~2 Lagrange point, at a
distance of approximately $1.5\times10^6\,\mathrm{km}$ from the Earth.
Due to the lack of Earth's geomagnetic shielding the satellite and the
detectors will be subject to cosmic ray impacts, modulated in
intensity by the solar cycle. The IXO orbit allows to point the
spacecraft in such a way, that the FOV is always facing away from the
Sun, thus theoretically allowing continuous observations. To
characterize the radiation background at L~2 we rely on model
estimates for the cosmic ray flux for our simulation. We use the
CREME96 model \cite{CREME96} with a fixed distance of $1.5\times
10^6\,\mathrm{km}$ above Earth for the planned launch date in 2020.
This date is near the solar cycle minimum, corresponding to a cosmic
ray flux maximum. Furthermore, we concentrate on the proton
contribution of the total cosmic ray flux, which is by far the most
dominant component. According to \cite{CREME96} the CREME model is
valid out to Mars orbit, which is at a distance from the Sun well
beyond L~2.

Fig.~\ref{fig:cremeComp} shows a comparison between the cosmic ray
proton spectrum calculated from different CREME models for both
missions Simbol-X and IXO and different launch dates. It is apparent
that in contrast to the older CREME86 model~\cite{CREME86}, the
CREME96 model gives a larger flux variation due to the influence of
the solar cycle. Please note, that the Simbol-X launch date of 2013 is
near the solar maximum (cosmic ray minimum) and the planned IXO launch
date of 2020 is close the solar minimum (cosmic ray maximum). The
satellite orbital position seems to have a negligible effect on the
resulting proton spectrum.

\section{Geant4 Simulations}
\begin{figure}
  \centering
  \includegraphics[width=0.99\columnwidth]{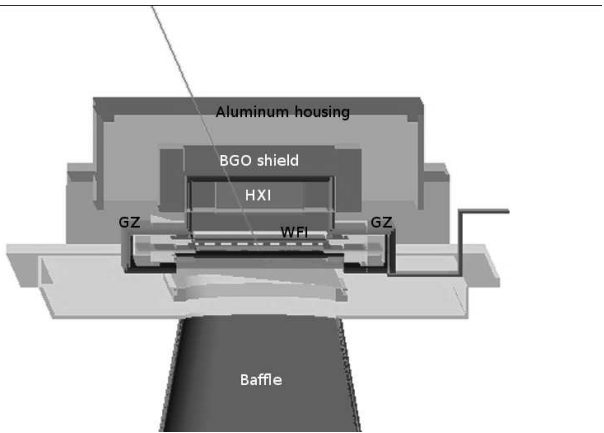}
  \caption{The Geant4 baseline geometry of the WFI and HXI
    detectors. The following components are shown from the outer to
    the inner: the aluminum housing, the BGO shield, the graded-Z
    shield, the HXI and the macro-pixel detector of the
    WFI.\label{fig:geometry}}
\end{figure}

The actual background simulations for IXO were done with the Geant4
Monte-Carlo software environment developed at CERN. Similar to
Simbol-X, we transfered the IXO detector geometry from the baseline
mechanical engineering model, abstracting some components in the
process in order to reduce computing time to a necessary minimum. The
Si wafer of the WFI and surrounding read-out electronics were modelled
with greatest detail, while the level of detail was reduced for more
distant geometry components. A graded-Z shield consisting of layers of
tantalum, tin, copper, aluminum and carbon was included in the model
as the innermost layers close to the wafer. The satellite structure
was modelled assuming a simplified geometry representation of the
movable and fixed instrument platform as well as the Sun shield. This
baseline geometry, without the satellite structures, is shown in
Fig.~\ref{fig:geometry} and will be used as a basis for further design
iterations and optimizations aimed at reducing the detector particle
background.

Our simulations for Simbol-X and IXO are based on the same standard
electromagnetic and low energy electromagnetic~\cite{geant4:lowEn}
Geant4 models, as well as a full set of hadronic physics which have
already been used for our Simbol-X simulations \cite{hauf:09a}. The
simulation of activation and radioactive decay processes is
optional. Our current simulations were done using Geant~4.9.2 p01.

Additional background reduction can be realized during data
post-processing by analysing pixel patterns and energy deposition of
events in the detector. An event is only considered as valid if it
meets certain criteria: the pixel pattern and the energy distribution
attributed to the event must fit into a specified valid pattern
mask. Furthermore, the deposited energy must be below a minimum
ionizing particle (MIP) threshold currently set to $15\,\mathrm{keV}$,
which is the maximum of the WFI energy range. For the case that an
invalid event pattern was registered in one read-out frame, we have
investigated the efficiency of different algorithms to reject the
event pattern: discarding only the affected pixels or the complete
frame in which the event was included. The discarding of whole frames
approximately halves the background rate but also introduces a dead
time of around $50\%$. Due to this only single patterns will be
discarded in future simulations.  For test purposes we have also
included a simplified geometric representation of the XMS experiment,
a microcalorimeter spectrometer, in our model, in order to study it's
influence on the WFI background.

\begin{table}[!t]
  \renewcommand{\arraystretch}{1.3} \centering
  \caption{Background estimates for Simbol-X and IXO.\label{tab:Simbolx-IXO-Background-Results}}
  \begin{tabular}{lrr}
    \hline
    & \multicolumn{1}{c}{Simbol-X}           & \multicolumn{1}{c}{IXO}\\ \hline
    Readout rate                             & $4000\,\mathrm{Hz}$ & $400\,\mathrm{Hz}$ \\ 
    Anti-coincidence                         & yes      & no      \\ 
    Raw count rate$^1$                       & $175$    & $ 2800$ \\ 
    Count rate after AC$^1$                  & $5.25 \pm 0.88$   &  NA     \\
    Count rate after pattern analysis$^1$    & $2.00 \pm 0.55$   & $77 \pm 3.7$    \\ 
    AC induced dead time                     & $18.7\%$ & NA      \\
    Efficiency AC                            & $97\%$   & NA      \\ 
    Efficiency pattern analysis.             & $62\%$   & $97\%$  \\\hline
  \end{tabular}

  \vspace{0.5em}
  {$^1$ \footnotesize count rates are given in units of $10^{-4}
    \mathrm{cts}\,\mathrm{cm}^{-2}\,\mathrm{s}^{-1}\,\mathrm{keV}^{-1}$}
\end{table}

\section{Simulation Results: Prompt proton induced background}
\begin{figure}[!t]
  \centering
  \includegraphics[width=0.99\columnwidth]{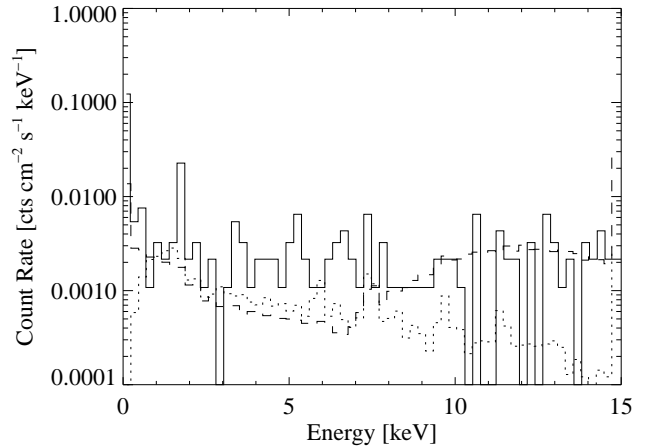}
  \caption{Simulated IXO WFI differential background spectrum (solid
    line) in comparison to the measured blank sky background of the
    Suzaku back (dashed line) and front illuminated CCDs of the XIS
    detector (dotted line)~\cite{suzaku:blanksky}.\label{fig:suzaku}}
\end{figure}
Our simulations for Simbol-X yield a count rate of $(2.0\pm 0.6)
\times 10^{-4}
\mathrm{cts}\,\mathrm{cm}^{-2}\,\mathrm{s}^{-1}\,\mathrm{keV}^{-1}$
for the LED with $18.7\%$ down time and $(2.6\pm 0.3) \times 10^{-4}
\mathrm{cts}\,\mathrm{cm}^{-2}\,\mathrm{s}^{-1}\,\mathrm{keV}^{-1}$
for the HED \cite{tenzer:09a}. The countrates given are with proper
anti-coincidence treatment and pattern analysis applied and are well
within the envisioned rates.

For IXO our simulations yield a preliminary estimate of the WFI background in
the $10^{-3}\,\mathrm{cts}\,\mathrm{cm}^{-2}\,\mathrm{s}^{-1}\,\mathrm{keV}^{-1}$
range for the baseline geometry. This background level is one order of
magnitude above the envisioned rate of
$10^{-4}\,\mathrm{cts}\,\mathrm{s}^{-1}\,\mathrm{keV}^{-1}$ and
consistent with background rates observed by currently flying missions
like Suzaku as shown in Fig.~\ref{fig:suzaku}. At the present level of
detail and statistics, the influence of the XMS on the WFI background
is negligible.

\begin{figure}[!t]
  \centering
  \includegraphics[width=1.01\columnwidth]{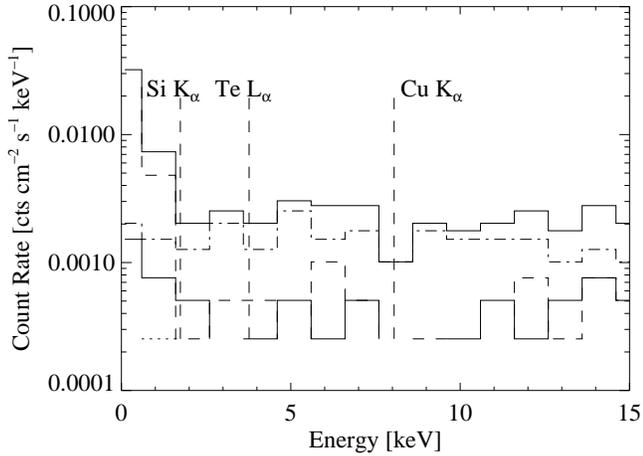}
  \caption{Simulated differential background spectrum of the WFI
    detector after pattern and MIP analysis. From top to bottom the
    following spectra are shown: the total background spectrum (solid
    line) and the contribution of electrons (dash-dotted line), gammas
    (solid line) and protons (dashed line) to the total background
    spectrum. For a detailed description see
    text. \label{fig:spectrum}}
\end{figure}

Our results show that the pattern and MIP detection algorithms used
can reliably reject $96\%$ of the background as invalid patterns. The
remaining $4\%$ of the overall background mainly originates from
secondary electron and primary proton energy depositions in the WFI
silicon chip as shown in the background spectrum in
Fig.~\ref{fig:spectrum}. Of these valid event patterns $74\%$ are
single pixel events, $24\%$ are double pixel events and a remaining
fraction of $2\%$ are triple pixel events. While events with $n>3$
dominate the raw background rate, they either have invalid pattern
shapes or deposit an energy which is above the MIP threshold or
commonly both. Furthermore, we observe a reduction of the background
of approximately 50\% in the case that we discard a complete frame if a
invalid event pattern is observed in this frame. Though this roughly
halves the background rate it also introduces a dead time of $>50\%$.

The WFI background spectrum shown in Fig.~\ref{fig:spectrum} also
demonstrates that the actual design of the graded-Z shield effectively
reduces any emission lines. Since electrons are the most prominent source of
the remaining background, which are not detectable through the applied
pattern or MIP rejection algorithms, future optimisation of the
mechanical design will focus on this issue.

\section{Progress of the Validation of Geant4 Radioactive Decay Simulations}
Our previous simulations for the Simbol-X focal plane detector module
yield an increase of the HED mean differential background flux due to
activation by cosmic ray protons from $\mathrm{2.6}\times
10^{-4}\,\mathrm{cts}\,\mathrm{cm}^{-2}\,\mathrm{s}^{-1}\,\mathrm{keV}^{-1}$
to $3.34 \times
10^{-4}\,\mathrm{cts}\,\mathrm{cm}^{-2}\,\mathrm{s}^{-1}\,\mathrm{keV}^{-1}$
\cite{tenzer:06a,chipaux:08a,tenzer:09a, hauf:09a}. Since we expect a
larger incident cosmic ray proton flux for the IXO mission time window
and orbit in comparison to the Simbol-X mission, we consequently
assume that the proton induced prompt and delayed background due to
activation will contribute with a similar or even larger amount to the
total detector background of the WFI. This assumption requires that we
have an accurate treatment of this background component in our
simulations.  Experience with existing Geant extensions like MGGPOD
(Geant3) and Cosima (Geant4) \cite{ weidenspointner:04a,
  Zoglauer05a:MEGAlib} further supports this assumption.

\begin{figure}
  \centering
  \includegraphics[width=0.99\columnwidth]{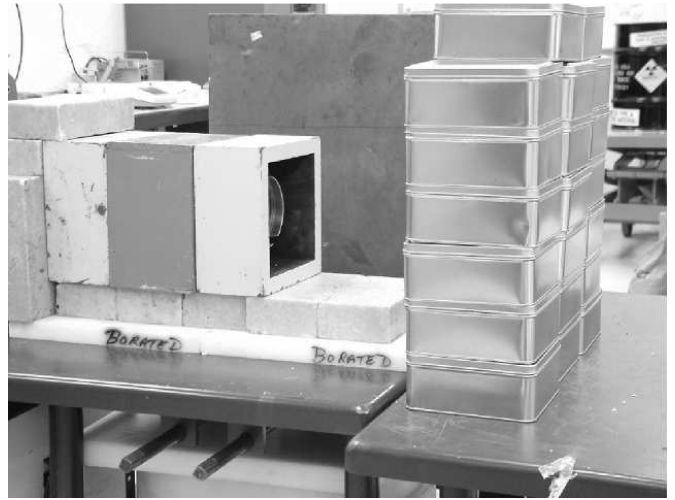}
  \caption{Experimental setup to measure the decay spectra of
    different isotopes (see text). On the left hand side the germanium
    spectrometer covered by a lead shielding is shown. The radioactive
    isotopes where placed into the gap between the spectrometer and the
    metal boxes on the left. \label{fig:radDecayGeo}}
\end{figure}

Because data on a systematic experimental verification of the
radioactive decay physics implemented in Geant4 has been rare, we have
started an experimental validation of the radioactive decay physics as
part of the Nano5 project~\cite{pia:09a}. In a first simple and
straight forward approach, we tried to reproduce measured spectra of
different radioactive sources with Geant4. The isotopes we used were
$^{137}$Cs, $^ {22}$Na, $^{54}$Mn, $^{60}$Co, $^{57}$Co and
$^{133}$Ba, with a specified activity of $37\,\mathrm{kBq}$ in June
2006. The decay spectrum of each individual isotope was observed with
an ORTEC GEM70P4 high purity Germanium detector with a $500
\mathrm{\mu m}$ thick Beryllium entrance window~\cite{ortec:gem}. The
detector provides an energy resolution of $1\,\mathrm{keV}$ at
$122\,\mathrm{keV}$ and $2\,\mathrm{keV}$ at $1.33\,\mathrm{MeV}$ and
was covered by a pair of copper and tin tubes and additional lead
shielding in order to suppress environmental gamma ray induced
background. The sources were placed at a known distance in a gap
between the detector and shielding components as shown in
Fig.~\ref{fig:radDecayGeo}. A background measurement was conducted
before and after each source measurement. The experimental spectra
were subsequently background subtracted and binned into
$1\,\mathrm{keV}$ energy intervals.

The geometry of the experimental setup, including all shielding
elements, and detector components was implemented as a Geant4 geometry
following the information provided by ORTEC [priv. comm.]. For our
simulations we induced $10^6$ decays for each isotope using the same
electromagnetic and hadronic physics as for our Simbol-X and IXO
background simulations. The source was modelled to emit to a solid
angle of $4\pi$. The simulated spectra were binned in the same way as
the measured spectra and finally normalized to the isotope's calculated activity on the measurement date, using an activity of $37\,\mathrm{kBq}$ in June 2006 as a reference point. The detector energy resolution was
approximated by folding the simulated data with a Gaussian function.

\begin{figure}[!t]
  \centering
  \includegraphics[width=0.99\columnwidth]{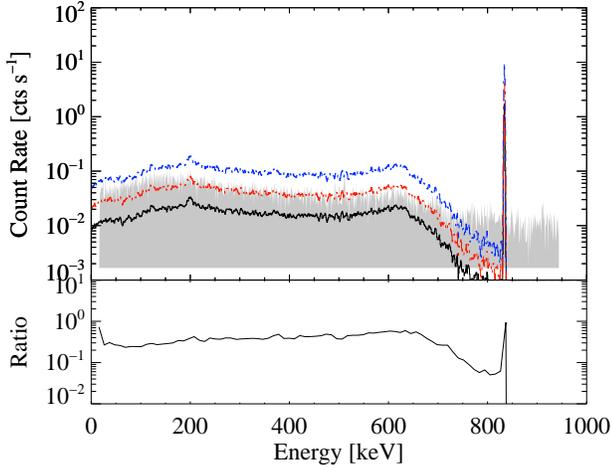}
  \caption{Experimentally measured and background subtracted $^{54}$Mn
    spectrum (grey shaded region) in comparison to simulated
    spectra. The simulated spectrum has been normalized to the
    activity of the source (bottom line, black), the continuum (middle line, orange)
    and the peaks (top line, blue). Notice the difference in the peak to
    continuum ratio. \label{fig:Mn}}
\end{figure}

\begin{figure}[!t]
  \centering
  \includegraphics[width=0.99\columnwidth]{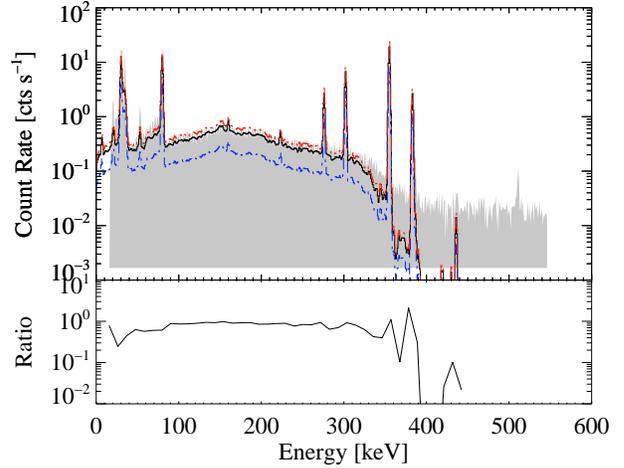}
  \caption{Experimentally measured and background subtracted
    $^{133}$Ba spectrum (grey shaded region) in comparison to
    simulated spectra. The simulated spectrum has been normalized to
    the activity of the source (middle line, black), the continuum (top
    line, orange) and the peaks (bottom line, blue). Notice the difference in peak to
    peak ratios.\label{fig:Ba}}
\end{figure}

Two examples of our results, a comparison of measured and simulated
spectra of two isotopes, $^{54}$Mn and $^{133}$Ba, are shown in
Figs.~\ref{fig:Mn}, \ref{fig:Ba}. It is obvious from
Figs.~\ref{fig:Mn}, \ref{fig:Ba} that the simulation is able to
qualitatively reproduce most of the spectral features (continuum shape
and emission lines). On the other hand there is a clear disagreement
between peak to peak and peak to continuum ratios by a factor of up to
$3$ between the simulated and measured spectra. In case of $^{54}$Mn the mean ratio of simulated data to experimental data is $0.3$ for the continuum compared to $0.9$ for the peak. Our comparision for $^{133}$Ba shows peak ratios varying between $0.6$ and $1.0$. This disagreement has
been observed for all measured isotopes, at different levels.

While there remains a systematic uncertainty in our flux normalisation
of the simulated data due to small uncertainties of the detector to
source distance or of the activity of the sources which affect the
overall normalization of the measured spectra, such a
disagreement of the peak to peak and peak to continuum ratios could be
of more serious nature and should be further investigated. Currently
we are focusing on two possibilities: either our Geant4 model is
over-simplified and we are missing important geometry parts or there
is an underlying problem in Geant4 physics.

Further measurements are currently in progress to investigate if this
problem exists for a broader variety of isotopes. At the same time we
investigate the influence of different geometrical parameters in the
simulation in order to quantify their influence on the result. Along
with analysing the contributions of the individual physics processes
involved this will lead to a better understanding of the origin of the
observed discrepancies.

\section{Conclusions and Outlook}
Our background estimates for the Simbol-X LED detector have
demonstrated that background rates in the
$10^{-4}\,\mathrm{cts}\,\mathrm{cm}^{-2}\,\mathrm{s}^{-1}\,\mathrm{keV}^{-1}$
range are achievable with the DEPFET detector technology. Preliminary
simulation results for the IXO WFI detector yield a background level
which is significantly larger compared to the anticipated science goal
and consequently the present baseline WFI design requires further
optimisation.

%% Our efforts concerning the experimental verification of Geant4
%% radioactive decay simulation have revealed a potential problem in
%% Geant4 physics. More data and more detailed systematic studies are
%% necessary to rule out any experimental effect.

In parallel to this work we are currently undertaking an experiment
with laser accelerated protons with the goal to measure the proton
induced activation and decay spectra of different graded-Z shield
materials which will be used for the IXO WFI.

\section*{Acknowledgement}
This work is supported by the Bundesministerium f\"ur Wirtschaft und
Technologie and the Deutsches Zentrum f\"ur Luft- und Raumfahrt - DLR
under the grant number 50QR0902.

\bibliographystyle{IEEEtran}
\bibliography{IEEEabrv,mnemonic,aa_abbrv,spacebgrdsim,conferences}

\end{document}